\documentclass{PoS}

\title{What do we understand  from multi-frequency monitoring of microquasars?}

\ShortTitle{Multi-frequency monitoring of microquasars}

\author{\speaker{Sergei A. Trushkin}\\
	Special Astrophysical Observatory RAS, Nizhnij Arkhyz, 369167, Russia\\
	E-mail: \email{satr@sao.ru}}
\author{Nikolaj N. Bursov \\
	Special Astrophysical Observatory RAS, Nizhnij Arkhyz, 369167, Russia\\
	E-mail: \email{nnb@sao.ru}}
\author{Nikolaj A. Nizhelskij \\
	Special Astrophysical Observatory RAS, Nizhnij Arkhyz, 369167, Russia\\
	E-mail: \email{nizh@sao.ru}}
\author{Elena  K. Majorova \\
	Special Astrophysical Observatory RAS, Nizhnij Arkhyz, 369167, Russia\\
	E-mail: \email{len@sao.ru}}
\author{Peter A. Voitsik \\
	Moscow State University, Physical department, Moscow, Russia\\
	E-mail: \email{voitsik@gmail.com}}

\abstract{
We discuss the results of the monitoring programs
of the X-ray binaries with relativistic jets studies.
We carried out a multi-frequency (1-30 GHz) daily monitoring of the
radio flux variability of the microquasars SS433, GRS1915+105, V4641 Sgr and
Cyg X-3 with RATAN-600 radio telescope during the recent sets in 2002-2006.
We detected a lot of bright short-time flares from GRS~1915+105 which could
be associated with active X-ray events.
In 2004 we have detected two flares from V4641 Sgr,
which followed after recurrent X-ray activity of the transient.
From September 2005 to May 2006 and then in July we have daily measured
flux densities from Cyg X-3.
In January 2006 we detected a drop down of its quiescent fluxes
(from 100 to $\sim$20 mJy), then the 1~Jy-flare was detected on 2 February
2006 after 18 days of quenched radio emission. The daily spectra
of the flare in the maximum were flat from 2 to 110 GHz, using the
quasi-simultaneous observations at 110 GHz with the RT45m telescope and
the NMA millimeter array of the Nobeyama Radio Observatory in Japan.
Several bright radio flaring events (1-15 Jy) followed during the continuing
state of very variable and intensive 1-12 keV X-ray emission ($\sim$0.5 Crab),
which was monitored in the RXTE ASM program.  We discuss the various
spectral and temporal characteristics of the light curves from the
microquasars. Thus we conclude that monitoring of the flaring radio
emission is a good tracer of jet activity X-ray binaries.}

\FullConference{VI Microquasar Workshop: Microquasars and Beyond\\
		September 18-22 2006\\
		Societ\`a del Casino, Como, Italy}

\begin{document}

\section{Introduction}
The first idea about cosmic radio sources variability belongs
to Shklovskij, who predicted the secular decreasing of the total radio
flux from the SNR Cas A \cite{shk60}. The remarkable simple formula related
a relative flux decrease because an adiabatic expansion with time:
$\Delta S_\nu/S_\nu=-2\gamma\Delta T/T$,
where $\gamma$ is a energetic spectral index, T -- an age of a source,
$\Delta T$ is a time interval between flux measurements.
The estimate gives 1.5\% per year for Cas~A, and it is not so far from
the measured value of ~0.9\% per year. Indeed,  few of improvements for such
evolution model of the radio remnant, included the additional acceleration of
the relativistic electrons, the complicated magnetic field evolution,
a slowing-down of the expansion can lead to a better estimate.

Thus, young Galactic SNRs were the first variable sources, but
real variability of the cosmic sources was detected from extragalactic
sources, quasars and AGNs, in sixties of last century \cite{sho62}.
Kardashev \cite{kar62} showed that a temporal variability is followed by a spectral
evolution of the radio sources, that could be used for a estimate of
the source age. The remarkable Slysh's formula \cite{sl63} gives us a way to
determine a magnetic field and therefore total energy, reserved in the
relativistic protons and electrons and magnetic fields.
Van der Laan developed Shklovskij's idea to generalize the main
formulae and showed that any synchrotron emitting source should evolve in
a similar manner \cite{laan66}.

Many long-ti\-me monitoring programs were begun: U. Michigan RAO \cite{AA},
GBI (NRAO/NASA) \cite{fiedler87a}, Mets\"ahovi \cite{ter92},
Bologna \cite{FFF83} and RATAN \cite{kov97} in the early 60-70ths.
They showed that almost all sources are variable, some on the year-decade
scale, others -- on the week-month scale.
The fast flickering of the sources was related to the ISM propagation
\cite{Hee84}.

Marscher and Gear \cite{mg85} were the first who to use Rees's idea
\cite{rees78} about internal shocks, running in the jets of flaring quasar
3C273.

An effective source of variable synchrotron emission in microquasars,
quasars and AGNs is outflows of accreting matter in the narrow cone --
the two-side relativistic jets, ejected from polar regions of accretion
disks around black holes or neutron stars.
The distinct clouds or plasmons contain magnetic fields and energetic
electrons.
The temporal and frequency dependencies in the light curves are a key
for clear understanding and good probe test for models of the physical
processes in cosmic jets. A comparison of the radio and X-ray variable emission
allows us to provide detailed studies.

\begin{figure}
\centering
\includegraphics[width=0.75\linewidth]{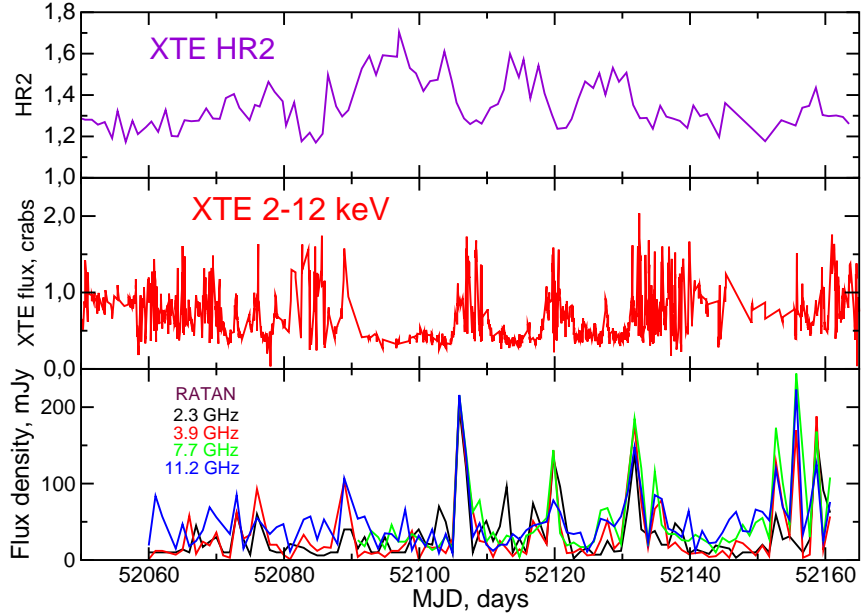}
\caption{Hardness ratio and light curves of GRS1915+105 at radio frequencies and at 2-12 keV
in 2001}
\label{fig:1}
\end{figure}

\section{Observations}

During each of the last five years we carried out observations 150-200 days
of with the RATAN-600 radio telescope for mQSO studies.
We have carried out the 250-day almost daily monitoring observations
of the microquasars Cyg X-3, GRS~1915+10, SS433, with the RATAN-600 radio
telescope at 1-22 GHz from September 2005 to May 2006.
The measured multi-frequency light curves can be directly compared with time
series of the X-ray observatory RXTE \cite{lev96}.

We have used a standard continuum radiometer complex. The
receivers at 3.9 (later 4.8), 7.7, 11.2, and 21.7 GHz were equipped
with closed-cycle cryogenic systems, which lowered the temperature of the
first amplifiers (HEMT) to 15 K. Low-noise transistor amplifiers were
installed in the 0.98-, 2.3-, and 30 GHz radiometers.
The observations were made using the `Northern sector' antenna of
the RATAN-600 radio telescope at the upper culmination.

The flux densities of the sources at four-six frequencies were measured in a
single observation of a source.
Although interference sometimes prevented realization of the maximum
sensitivity of the radiometers, daily observations of reference sources
indicate that the error in the flux density measurements did
not exceed 5\% at 1.0, 2.3, 3.9, and 11.2 GHz and 10-15\% at 21.7 and 30 GHz.

The standard flux error could be calculated from:
      $$ (\Delta S_\nu)^2 = (a_\nu)^2 + (b_\nu*S_\nu)^2 $$,
where $a_\nu$ is a radiometer noise component, and $b_\nu$ is an antenna
and calibration instability component. The values of the $a_\nu$ and $b_\nu$
are given in Table \ref{tab:1}.

\begin{table}
\begin{center}
\caption{Sensitivity parameters of the RATAN-600 telescope.}\vspace{1em}
\renewcommand{\arraystretch}{1.2}
\begin{tabular}{lllllllll}
\hline
 $\lambda$, cm & 31  & 13.0 &  ~7.6  & ~6.2  & ~3.9 &  ~2.7 & ~1.4  & ~1.0 \\
 $\nu$, GHz    & 1.0 & ~2.3 &  ~3.9  & ~4.8  & ~7.7 &  11.2 & 21.7  & 30.0 \\
   $a_\nu$,mJy & 30  & 10   &   3    &  3    &  5   &   5   &  10   & 15   \\
   $ b_\nu$    & 0.01& 0.01 &   0.02 &  0.015& 0.02 &  0.03 & 0.05  & 0.04 \\
\hline
\end{tabular}
\label{tab:1}
\end{center}
\end{table}

The flux density calibration was performed using observations of 3C286
(1328+30), PKS 1345+12 and NGC7027 (2105+42). We have controlled the antenna
gain with a thermal source (HII region) 1850--00 in daily observations also.
We accepted the fluxes for these sources from our calibrator list \cite{ali85},
which, in turn, were consistent with the primary radio astronomy flux scale
from \cite{baars77} and with the new flux measurements from \cite{ott94}.

\section{Discussion}
\subsection{GRS 1915+105: X-ray -- radio correlation}

The X-ray transient source GRS~1915+105 was discovered in 1992 by
Castro-Tirado et al. \cite{CT92} with the WATCH instrument on board GRANAT.
An apparent super-luminous motion of the jet components was detected
and  the determination of 'microquasars' were coined \cite{mr94}.
Fender et al. \cite{fender02} discussed the alert observations of  two
flares (July 2000), when for the first time detected the quasi-periodical
oscillations with P = 30.87 minutes at two frequencies: 4800 and 8640 MHz.
Linear polarization was measured at a level 1-2 per cent
at both frequencies as well.

\begin{figure}
\centering
\includegraphics[width=0.75\linewidth]{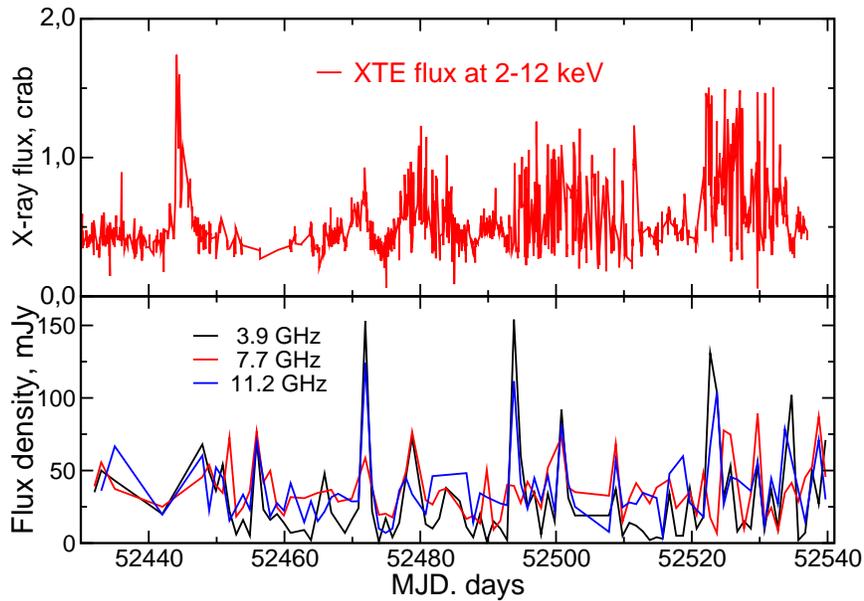}
\caption{Light curves of GRS1915+105 at radio frequencies and at 2-12 keV
in 2002}
\label{fig:2}
\end{figure}

In Fig.\ref{fig:1} and \ref{fig:2} the radio and X-ray light curves are
showed during the sets in 2001 and 2002. The most bright radio events have
some associated events from the X-ray light curves
received in ASM RXTE at 2-12 keV.

\begin{figure}
\centering
\includegraphics[width=0.8\linewidth]{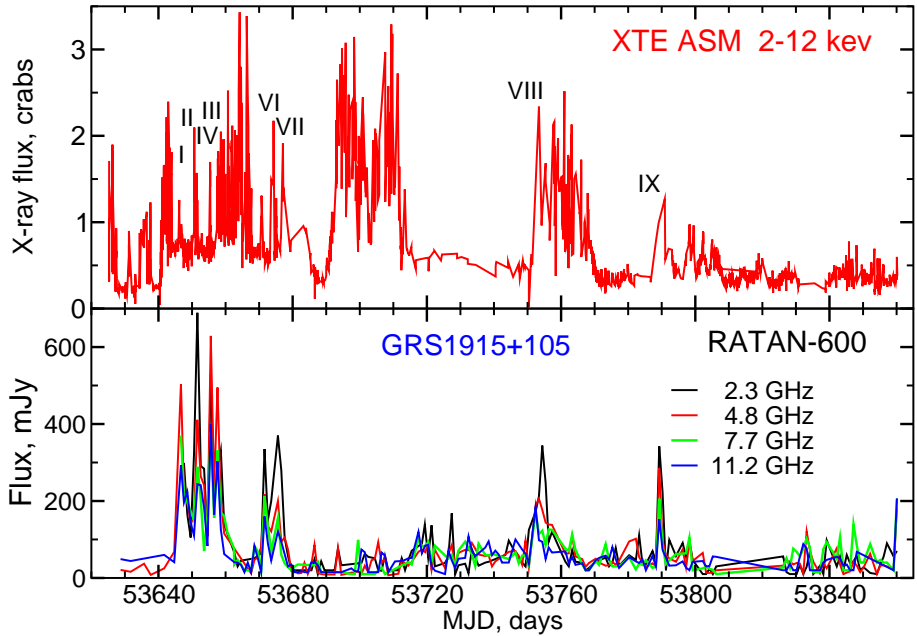}
\caption{Light curves of GRS1915+105 at radio frequencies and at 2-12 keV
from September 2005 to March 2006.}
\label{fig:3}
\end{figure}

In Fig.\ref{fig:3} the radio and X-ray light curves are showed for the
total set. The nine radio flares have the counterparts in X rays.
The radio spectrum was optically thin in the first two flares,
and optically thick in the third one (see details in \cite{nam07}).

The profiles of the X-ray spikes during the radio flares
are clearly distinguishable from other spikes because its shape
shows a fast-rise and a exponential-decay. The X-ray spikes,
which reflect activity of the accretion disk, show an irregular
pattern. During a bright radio flare, the spectra of the X-ray spikes become
softer than those of the quiescent phase, by a fraction of $\sim$30\%
\cite{nam07}.

\begin{figure}
\centering
\includegraphics[width=0.85\linewidth]{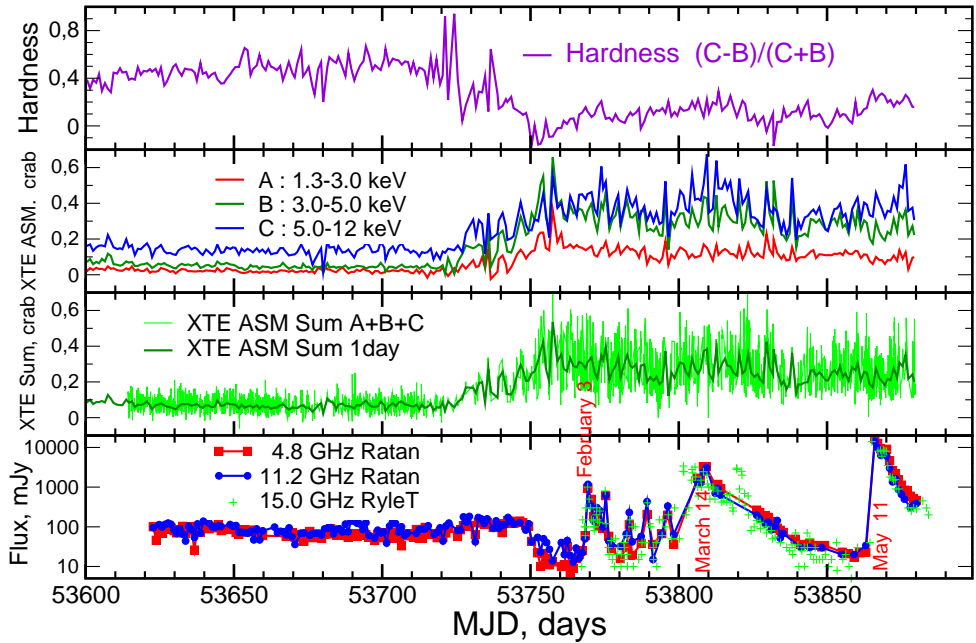}
\caption{
The RATAN and RXTE ASM light curves of Cyg X-3 from September 2005 to
May 2006.}
\label{fig:4}
\end{figure}
\begin{figure}
\centering
\includegraphics[width=0.73\linewidth]{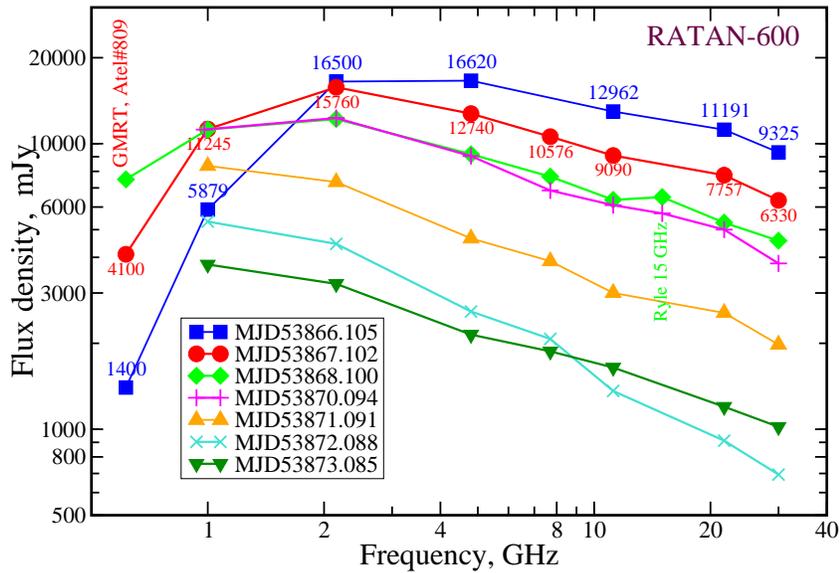}
\caption{The daily spectra of Cyg X-3 during the flare in May 2006.}
\label{fig:5}
\end{figure}

Miller at al. \cite{miller06} have detected a one-side large-scale radio jet
with VLBA mapping during an X-ray and (XI) radio outburst on 23 February
2006 (MJD53789.258).  Then the optically thin flare with fluxes
340, 340, 342, 285, 206, and 153 mJy was detected at frequencies
1, 2.3, 4.8, 7.7, 11,2 and 21.7 GHz.

\subsection{Cyg X-3: 2006 -- new long-term active period }

During ~100 days (September -- December 2005)
Cyg~X-3  was in a quiescent state of $\sim$100 mJy (Fig.\ref{fig:4}).
In December 2005 its X-ray flux began to increase and the radio flux at 2-11
GHz increased also. Then the flux density of the source at 4.8 GHz  was
found to drop from 103 mJy on 2006 Jan 14.4 (UT) to 43 mJy on Jan 15.4 (UT),
and to 22 mJy on Jan 17.4 (UT). The source is known to exhibit radio
flares typically with a few peaks exceeding 1-5 Jy following such a quenched
state as Waltman et al. \cite{waltman94} have showed in
the intensive monitoring of Cyg X-3 with the Green Bank
Interferometer at 2.25 and 8.3 GHz.
The source has been monitored from 2006 Jan 25 (UT) with the Nobeyama Radio
Observatory 45m Telescope (NRO45m Telescope),
the Nobeyama Millimeter Array (NMA), Yamaguchi-University 32-m
Radio Telescope (YRT32m), and Japanese VLBI Network telescopes.
On Feb 2.2 (UT), about 18 days after it entered the quenched state, the
rise of a first peak is detected with the NRO45m Telescope and YRT32m.
On Feb 3.2 (UT), the flux densities reached the first peak at all the
sampling frequencies from 2.25 GHz to 110.10 GHz (\cite{tsuboi06}).
The spectrum at maximum (3 February) of the flare was flat
as measured by RATAN, NRO RT45m and NMA from 2 to 110 GHz.
The next peak of the active events on 10 February
reached the level of near 1 Jy again with a similar flat spectrum.
Then three short-time flares have happened during a week.
The flare on 18 February had the inverted spectrum with the same
spectral index $\alpha$=+0.75 from 2.3 to 22 GHz.

\begin{figure}
\centering
\includegraphics[width=0.7\linewidth]{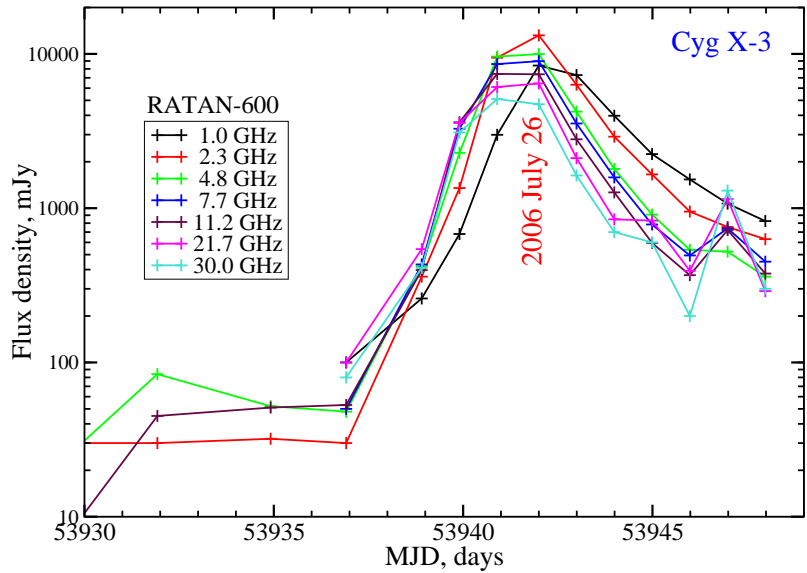}
\caption{
The RATAN and RXTE ASM light curves of Cyg X-3 in July 2006}
\label{fig:6}
\end{figure}

\begin{figure}
\centering
\includegraphics[width=0.7\linewidth]{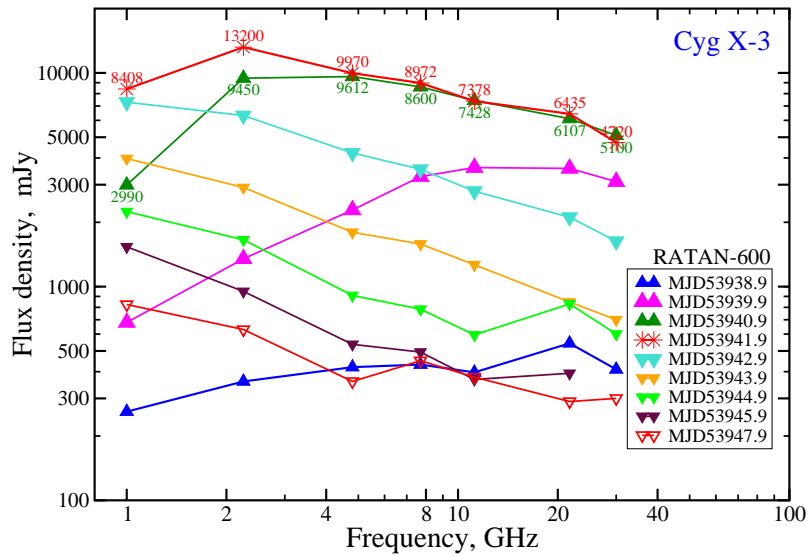}
\caption{The daily spectra of Cyg X-3 during flare in July 2006.}
\label{fig:7}
\end{figure}
In the active period there were two powerful flares, March 14 to 3-5 Jy
and May 11 to 12-16 Jy at 2-30 GHz. In the May flare
fluxes have grown up by a factor $\sim$1000 during a day.
Such powerful ejection of relativistic electrons
were detected with RATAN and by G. Pooley with the Ryle telescope.

The change of the spectrum  during the flare on May 11-19
followed the model of a single ejection of the relativistic electrons, moving
in thermal matter in the intensive WR-star wind.
It stays in optically thin mode at  higher frequencies,
meanwhile at lower frequency 614 MHz (\cite{pal06}, Fig.\ref{fig:5}).
Cyg X-3 was in hard absorption due to thermal electrons in its stellar wind.
In the continuing active state of Cyg X-3 we detected a very fast-rise flare
at 2.3 and 8.5 GHz with RT32 (IAA) on 5 June 2006 (MJD53891) \cite{tru06}.
During 3 hours the fluxes changed from $\sim1$ Jy to 2 Jy and then
decreased to 100-400 mJy during 15 hours.

In Fig.\ref{fig:6} the light curves of the July flare are shown.
We detected the total phase of the flux rising during 4-5 days. Thus
for the first time we could clearly see the evolution of the spectrum  during
the start phase of the flare (Fig.\ref{fig:7}). And it was amazing that
the low-frequency part of the spectra evolved from nearly flat optically
thin (at 1 GHz) on the first day to optically thick after 3-4 days.
Obviously, in the standard model of the expansion of the compact
sources (jets components) there is no explanation for such behaviour.
We had to conclude that the thermal electron density, responsible for the
low frequency absorption, grows up during grow-up of the relativistic electron
density.
The later stage of the spectral and temporal evolution could be
fitted by the modified finite segments model by Marti et al. \cite{marti92} or
Hjellming \cite{Hj88} and Hjellming et al. \cite{Hj00}.
Indeed in Fig.\ref{fig:7} the radio spectra of the July flare
evolved from the fourth day (MJD53942) as usual adiabatically expanding
relativistic jets with $v_{jet}\sim0.74c$.
Taking into account a boosting effect, the best fitted parameters
are: ejection time $t_i\sim 8.5$ days, exponential expansion stage
$t_{ex}\sim 7.5$ days,
thermal electron density at $T_e = 10^4$ K, $n_{th} = 2\,\,10^4 cm^{-3}$,
magnetic field  $B_0 = 0.07$ Gs, and enegry index $\gamma = -1.85$.
During the rising stage of the flare we should involve
the intense internal shocks running through the jet as proposed Wadanabe et
al. \cite{wad03}.

\begin{figure}
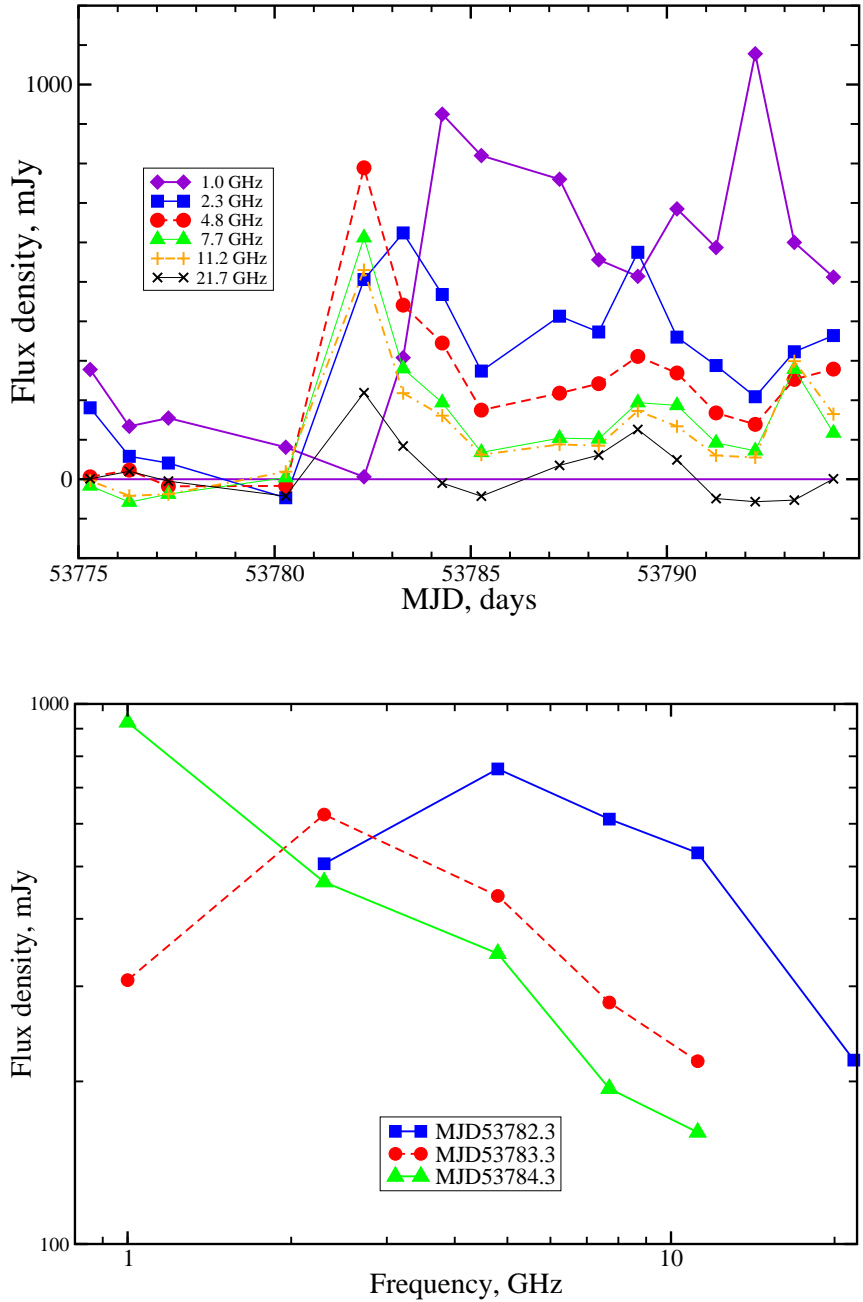

\centering
\includegraphics[width=0.75\linewidth]{ss433_f1.eps}

\vspace{1.cm}

\includegraphics[width=0.75\linewidth]{ss433_f2.eps}
\caption{The light curves after subtracting a quiet spectrum of SS433
in February 2006 and the flaring spectra  on 16-18 February 2006.}
\label{fig:8}
\end{figure}

\subsection{SS433: new flaring events and their spectra}

The first microquasar SS433, a bright variable emission star was
identified  with a rather bright compact radio
source 1909+048 located in the center of a supernova remnant W50.
In 1979 moving optical emission lines, Doppler-shifted due to precessing
mass outflows with 78000 km/s, were discovered in the spectrum of
SS433. At the same time in 1979 were discovered a unresolved
compact core and 1 arcsec long aligned jets in the MERLIN radio
image of SS433. Since 1979  many monitoring sets
(for ex., with GBI \cite{fiedler87b}, RATAN \cite{tru03}) were began.
Different data do indicate a presence of a very
narrow (about 1$^o$) collimated beam at least in X-ray and optical
ranges. At present there is no doubt that SS433 is related to W50.
A distance of near 5 kpc was later determined
by different ways including the direct measurement of proper motions of
the jet radio components.

Kotani et al. \cite{kotani06} detected the fast variation in the X-ray
emission of SS433 during the radio flares, and probably QPOs of 0.11 Hz.
Massive ejections during this active period could  be the reason of
such behavior.
The daily RATAN light curves are measured from September 2005 to May 2006.
The activity of SS433 began during the second half of the monitoring
set. Some flares happened just before and after the multi-band program
of the studies of SS433 in April 2006 \cite{kotani07}.

In the top of Fig.\ref{fig:8} the light curves during the bright flare in February 2006 are
showed after subtracting a quiet spectrum $S_\nu[Jy] =1.1\nu^{-0.6}[GHz]$.
The delay of the maximum flux at 1 GHz is about 2 days and 1 day at 2.3  GHz
relative to the maxima at higher frequencies.
Below in Fig.\ref{fig:8} the spectra of the flare during the first three days.
We see a characteristic shift of turn-over of spectra to low frequency
during the flare, clearly indicating a decrease of the absorption with time.

\section{Conclusions}
The RATAN microquasar monitoring data give us abundant material for
comparison with X-ray data from the ASM or ToO programs with RXTE, CHANDRA, Suzaku and
INTEGRAL. The 1-30 GHz emission originates often from different optically thin
and thick regions. That could give us a key for adequate modelling of the
flaring radio radiation formed in the relativistic jets interacting with
varying circumstellar medium or stellar winds.

{\it Acknowledgments}.
These studies are supported by the Russian  Foundation  Base Research (RFBR)
grant N~05-02-17556 and the mutual RFBR and
Japan Society for the Promotion of Science (JSPS) grant N~05-02-19710.

\end{document}